\documentclass[english]{gretsi}

\usepackage[utf8]{inputenc}

\usepackage[T1]{fontenc}

\usepackage{amsmath, amssymb, amsfonts, subcaption}

\newcommand{\eg}{\textit{e.g.},~}

\begin{document}


\titre{AutoMashup: Automatic Music Mashups Creation}

\auteurs{
  \auteur{Marine}{Delabaere}{delabaere.marine@gmail.com}{1,$\star$}
  \auteur{Léa}{Miqueu}{lea.miqueu@gmail.com}{1,$\star$}
  \auteur{Michael}{Moreno}{michael.moreno.st@gmail.com}{1,$\star$}
  \auteur{Gautier}{Bigois}{}{1}
  \auteur{Hoang}{Duong}{}{1}
  \auteur{Ella}{Fernandez}{}{1}
  \auteur{Flavie}{Manent}{}{1}
  \auteur{Maria}{Salgado-Herrera}{}{1}
  \auteur{Bastien}{Pasdeloup}{bastien.pasdeloup@imt-atlantique.fr}{1}
  \auteur{Nicolas}{Farrugia}{nicolas.farrugia@imt-atlantique.fr}{1}
  \auteur{Axel}{Marmoret}{axel.marmoret@imt-atlantique.fr}{1}
}

\affils{
   \affil{1}{IMT Atlantique, Lab-STICC, UMR CNRS 6285, F-29238, Brest, France}
   \affil{$\star$}{$\text{Equal contribution}$}

}


\resume{
Nous présentons AutoMashup, un système de création automatique de \textit{mashups} basé sur la séparation de sources, l’analyse musicale et l’estimation de compatibilité. Nous proposons d’utiliser COCOLA pour évaluer la compatibilité entre pistes séparées, et explorons dans quelle mesure des modèles audio généralistes préentraînés (CLAP et MERT) peuvent estimer cette compatibilité sans apprentissage supplémentaire. Nos résultats révèlent que la compatibilité est asymétrique --- elle dépend du rôle attribué à chaque piste (voix ou accompagnement) --- et que les modèles étudiés ne reproduisent pas la cohérence obtenue avec COCOLA. Ces observations soulignent les limites actuelles des représentations audio généralistes pour l’estimation de compatibilité musicale dans les mashups.}

\abstract{We introduce AutoMashup, a system for automatic mashup creation based on source separation, music analysis, and compatibility estimation. We propose using COCOLA to assess compatibility between separated stems and investigate whether general-purpose pretrained audio models (CLAP and MERT) can support zero-shot estimation of track pair compatibility. Our results show that mashup compatibility is asymmetric --- it depends on the role assigned to each track (vocals or accompaniment) --- and that current embeddings fail to reproduce the perceptual coherence measured by COCOLA. These findings underline the limitations of general-purpose audio representations for compatibility estimation in mashup creation.}

\maketitle


\section{Introduction}
Music mashups are musical works that combine elements --- typically vocal and instrumental stems --- from two or more pre-existing recordings to create a new, coherent composition. A common form involves overlaying the isolated vocals of one song onto the instrumental backing of another. Mashups have become increasingly popular in contemporary music culture, particularly in DJ performances and online communities~\cite{popenstock2024mouth}. However, producing high-quality mashups is far from trivial: it requires not only technical skills (\eg editing, beatmatching, pitch shifting) but also a deep understanding of musical compatibility --- how harmony, rhythm, and structure interact across songs~\cite{brovig2012contextual}. As a result, mashup creation remains largely inaccessible to beginners, despite the growing interest in creative music recombination.

Hence, several systems have been proposed to assist or automate mashup creation~\cite{davies2014automashupper, xing2020pomash, huang2021modeling, wu2024graph}. In this paper, we present ``AutoMashup,'' a new tool to automatically create mashups, using a modern pipeline that includes source separation, using Demucs~\cite{rouard2023hybrid}, and high-level musical information --- such as tonality, structure, beats --- using Allin1~\cite{kim2023all}. 

We also propose using a novel deep learning model to automatically select coherent song pairs for mashups: COCOLA~\cite{ciranni2025cocola}. COCOLA is designed for accompaniment compatibility, which closely aligns with our task and correlates well with perceptual evaluations\footnote{Empirically, we observed a strong correlation between COCOLA scores and subjective mashup quality.}. However, because COCOLA requires access to both songs for each evaluation, its use implies a combinatorial cost when searching across large song libraries. To address this, we investigate whether general-purpose audio models --- CLAP~\cite{wu2023large} and MERT~\cite{li2023mert} --- can support zero-shot compatibility estimation. While promising in principle, our results show that these models fail to capture the musical coherence required for effective mashups.




The remainder of this paper is structured as follows. Section~\ref{sec:related_work} reviews existing research on mashup generation and compatibility modeling. Section~\ref{sec:automashup} introduces AutoMashup, the mashup generation tool we use in this study. Section~\ref{sec:automatic_song_selection} presents our methodology for evaluating song selection using CLAP, MERT, and COCOLA scores. Section~\ref{sec:experiments} reports our experimental results, followed by a conclusion discussing current limitations and future research directions. Finally, all the code can be found on our companion GitHub repository\footnote{\url{https://github.com/ax-le/automashup}}.

\section{Related Work} 
\label{sec:related_work}
Several systems for automatic mashup creation have been proposed in recent years~\cite{davies2014automashupper, xing2020pomash, huang2021modeling, wu2024graph}, and many share design principles with AutoMashup. Most of these systems rely on a combination of tempo synchronization, key matching, and segment-level alignment --- for example, overlaying chorus sections or aligning verse boundaries --- when combining tracks \cite{davies2014automashupper, xing2020pomash}. These techniques aim to ensure what Brøvig-Hanssen and Harkins describe as musical congruence \cite{brovig2012contextual}, where rhythm and harmony are closely aligned to produce perceptually coherent results --- even when there is deliberate stylistic or semantic contrast between the original songs.

The question of how to automatically select compatible song pairs has also received attention. Early approaches, such as AutoMashUpper~\cite{davies2014automashupper}, relied on rule-based estimations of harmonic and rhythmic similarity using chroma features and beat tracking. PopMash \cite{xing2020pomash} extended this by incorporating melodic similarity and phonetic analysis of lyrics. More recent systems have begun exploring learning-based approaches to compatibility estimation, including contrastive learning frameworks that model pairwise stem compatibility \cite{huang2021modeling}, and graph neural networks that represent stems and their interrelations explicitly \cite{wu2024graph}.

Our work complements these contributions by testing whether pretrained audio models (CLAP~\cite{wu2023large} and MERT~\cite{li2023mert}) can support scalable, zero-shot song selection. We also explore the use of modern audio tools --- Demucs~\cite{rouard2023hybrid} for stem separation, Allin1~\cite{kim2023all} for structure and key estimation, and COCOLA~\cite{ciranni2025cocola} for compatibility evaluation --- which, to our knowledge, have not yet been combined in this context.

\section{Mashup creation with AutoMashup}
\label{sec:automashup}
AutoMashup combines several signal processing and deep learning methods to create mashups, which we present hereafter. In short, after separating the sources of both songs, our system estimates musical information such as key, music structure, tempo, and beats, which are then used to align both songs, both harmonically and rhythmically. AutoMashup also contains an API powered by \textit{Streamlit} and \textit{Barfi}. The entire AutoMashup system is open-source.

\subsection{Source Separation}

Source separation is the first stage, used to extract the vocal track from the rest. While its use in mashup creation is not new~\cite{huang2021modeling}, we leverage the recent Demucs model~\cite{rouard2023hybrid}, considered the state-of-the-art for this task.

\subsection{Estimating Musical Information}

AutoMashup leverages the Allin1~\cite{kim2023all} music analyzer to decompose songs into key components. Allin1 extracts various musical features, including tempo, beats and downbeats, key, and segments, which are different song sections (such as intros, verses, and choruses).

This detailed breakdown allows users to isolate specific parts of songs (vocals from the chorus, bass from the verse, drum solo, etc.) to be used in mashups, and allows for enhanced modification (pitch and key shifting, tempo modification, etc.).

\subsection{Aligning Songs} 
Starting from several songs (in general, two), AutoMashup takes the first song as the base song, aligning the others on the information of the base song (\eg tonality, duration, and BPM). Then, it aligns the segments (verse with verse, chorus with chorus, etc.) and modifies the tonality. 

When aligning the segments, if the duration of a segment in the base song is longer than the corresponding segment in the second song, AutoMashup repeats the second song’s segment to fill the gap. This process ensures that both songs remain synchronized throughout the mashup. However, one limitation of this method arises when a segment exists in the base song but is missing in the second song. In such cases, AutoMashup will continue playing only the segment from the base song. 
When aligning two songs, AutoMashup adjusts the BPM of the second song to match the tempo of the base song (both at the beat and downbeat scales). 
This ensures that both songs follow the same rhythmic structure, keeping their timing in sync and making the transitions between them smoother.

One of the essential challenges in creating mashups is adjusting the pitch and key of the individual tracks so that they harmonize. AutoMashup relies on Allin1’s key analysis and applies key adjustments to adjust the pitch where necessary. This step ensures that the combined elements are musically compatible, preventing dissonance in the final mashup. 

Key and BPM alignment are implemented using the library Pyrubberband~\cite{pyrubberband}.




\section{Automatic Song Selection}
\label{sec:automatic_song_selection}
As it is, AutoMashup asks the user to provide both songs to be combined. In this section, we will present a tentative system to automatically select songs to be combined. The experiments can be found on the companion GitHub repository.

\subsection{COCOLA score}

To quantitatively assess compatibility, we employ COCOLA~\cite{ciranni2025cocola}, a contrastive model trained to evaluate harmonic and rhythmic coherence between audio segments, leading to a COCOLA score of compatibility between two stems. While COCOLA was trained exclusively on instrumental music, we found that its scores empirically correlate with our own qualitative assessments of mashup quality --- even in the presence of vocals. Although a formal validation of this assumption is beyond the scope of this study, we identify it as an important direction for future work. To the best of our knowledge, this is the first study that leveraged COCOLA for automatic song selection for mashups.

COCOLA, while promising for evaluating musical compatibility, requires direct comparison between every possible track pair, resulting in computationally prohibitive $O(n^2)$ complexity for large music libraries. Hence, automatic song selection would benefit from a more scalable approach for assessing musical compatibility. 

\subsection{General-purpose Audio Models}

To address the computational complexity of COCOLA's pairwise track comparisons, we explore using large audio models such as CLAP~\cite{wu2023large} and MERT~\cite{li2023mert}. We hypothesize that the rich representational capabilities learned by these deep learning models can be leveraged to efficiently approximate track similarity. These models produce embeddings --- compressed, fixed-length vector representations of audio signals, extracted from their internal latent space --- that capture both semantic and acoustic characteristics. We investigate whether the embeddings obtained for each song can be used to predict mashup compatibility. Specifically, we hypothesize that songs with similar embeddings are more likely to blend well in a mashup, and that the distance between embeddings should correlate with their compatibility, as estimated by the COCOLA score. To test this, in Section~\ref{sec:experiments}, we analyze the cosine similarity between the vocal and instrumental stem embeddings for each candidate song pair.


\section{Experiments}
\label{sec:experiments}

\subsection{Dataset}
Due to the exponential cost of computing COCOLA scores for all song pairs, we subselected 21 tracks from the FMA dataset~\cite{defferrard2017fma}, guided by key annotations from Stella Wong’s repository~\cite{wong2023fma}. We intentionally limited diversity to simplify mashup creation and better focus on the specific characteristics of each track. Future work should expand this study to more diverse musical material.

To reduce structural complexity and ensure vocal content, we excluded experimental and jazz tracks, instrumental pieces, and classical music, which often involve complex arrangements or choirs. As a result, we limited our selection to popular music. Within this subset, we focused on songs within two semitones of C (B$\flat$ to D), as C major is the most common key in popular music~\cite{hooktheory2020analysis}, and to minimize the potential artifacts introduced by repitching in mashups. Only tracks in major keys were retained, given their dominance in popular genres~\cite{richards2017tonal}. We further constrained durations to the 184-194 second range, centered around the dataset’s mean (189s). This filtering process yielded 21 tracks, resulting in 420 music samples when combined with base tracks and their generated mashups.


\subsection{Asymmetry in the Song Selection}

When computing cosine similarities between (instrumental and vocal) track embeddings generated with CLAP and MERT, we found that the resulting similarity matrices were systematically asymmetric. Specifically, the similarity between the vocal track from song A and the instrumental track from song B differed from the reverse pairing (vocal of song B, instrumental of song A), despite involving the same musical content\footnote{We qualitatively confirmed this asymmetry by listening to a few mashups.}. This directional discrepancy indicates that musical compatibility is not symmetric, and depends heavily on the functional role assigned to each track (\eg vocals or accompaniment). In addition, it implies that mashup compatibility cannot be solely derived via musical features (such as tonality or tempo, as in~\cite{davies2014automashupper}), because such features are consistent between the vocal and instrumental tracks. We did not find explicit mention of asymmetry in previous works~\cite{davies2014automashupper, xing2020pomash, huang2021modeling}.

This finding had direct implications for mashup generation. It suggested that compatibility cannot be modeled as a symmetric relation, and that successful pairings depend on assigning consistent roles to the tracks being combined. 

\subsection{Clustering Analysis}
\label{sec:clustering}
\begin{figure*}
            \centering
            \hfill
            \begin{subfigure}{0.45\linewidth}
                \centering
                \includegraphics[width=0.8\linewidth]{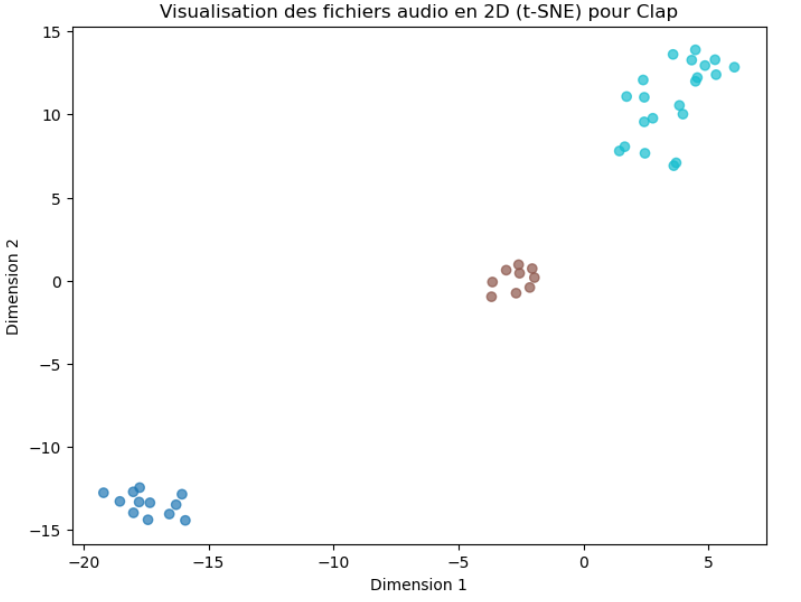}
                \caption{CLAP}
            \end{subfigure}
            \hfill
            \begin{subfigure}{0.45\linewidth}
                \centering
                \includegraphics[width=0.8\linewidth]{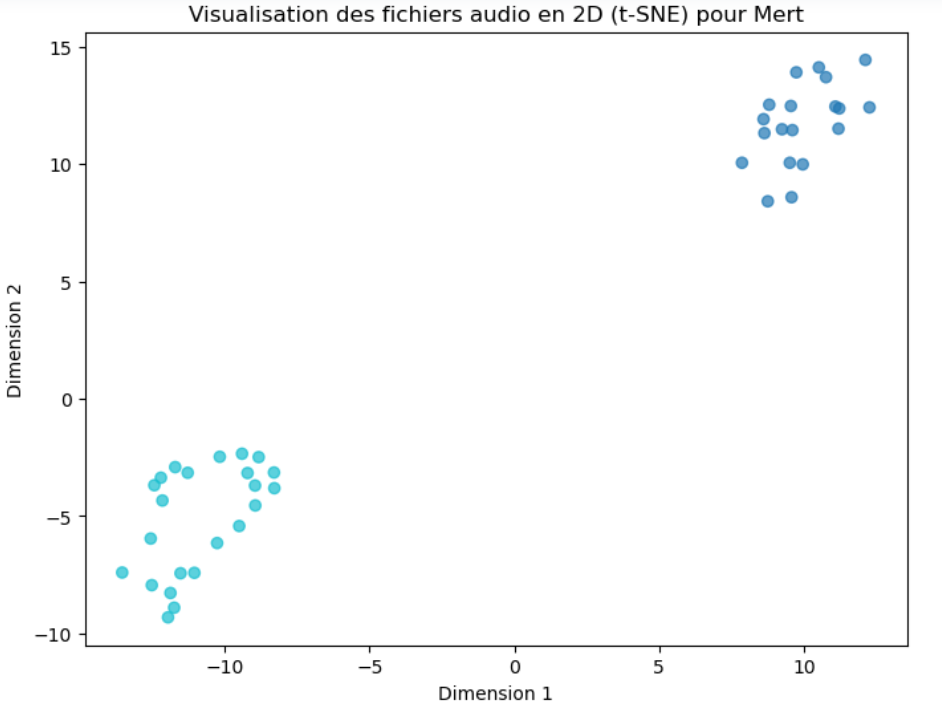}
                \caption{MERT}
            \end{subfigure}
            \hfill
            \caption{t-SNE visualisation of the embeddings obtained by CLAP and MERT. The colors represent different clusters, obtained via hierarchical clustering.}
            \label{fig:scatter}
        \end{figure*}

    To further investigate the embeddings produced by CLAP and MERT, we applied hierarchical clustering on 42 embeddings corresponding to the separated vocal and instrumental tracks of 21 songs. Clustering was performed by thresholding pairwise cosine distances, and visualized using t-SNE projections shown in Figure~\ref{fig:scatter}.

    The t-SNE visualizations reveal clear differences in the structure of the two embedding spaces. CLAP embeddings formed three distinct clusters: one dominated by vocal tracks, one by instrumental tracks, and a third cluster consisting mostly of synthetic voices. This separation suggests that CLAP, trained on language-audio pairs, organizes its latent space primarily along semantic and acoustic boundaries --- for instance, distinguishing human speech from instrumental content. However, this modality-based organization leads to a separation between vocals and instrumentals, which may limit CLAP's usefulness for modeling cross-stem compatibility in mashups.
    
    In contrast, MERT embeddings produced only two clusters, each containing a mix of vocal and instrumental tracks. This pattern suggests that MERT encodes information that cuts across the vocal/instrumental divide --- possibly focusing more on tonal, rhythmic, or structural properties. In the context of mashup creation, such an organization could reflect a grouping of musically compatible tracks, regardless of their stem type.
    
    To quantitatively compare the clustering outputs of CLAP and MERT, we computed the Adjusted Rand Index (ARI)~\cite{rand1971objective}, a standard measure of clustering similarity that corrects for chance. Our analysis yielded an ARI of -0.0261, indicating no agreement between the clusterings beyond random chance.
    

    This result implies that CLAP and MERT capture different aspects of musical content --- likely because MERT is music-specific, while CLAP is trained on broader audio-text data.

\subsection{Comparison of CLAP and MERT embeddings with COCOLA Score}
\label{sec:embed_cocola}

We conducted a correlation analysis to evaluate how well similarity scores derived from CLAP and MERT embeddings align with the COCOLA score, presented in Table~\ref{tab:correlation}.

\begin{table}[h]
    \centering
    \resizebox{\columnwidth}{!}{ 
    \begin{tabular}{lcc}
        \toprule
        \textbf{Correlation Metric} & \textbf{CLAP Embeddings} & \textbf{MERT Embeddings} \\
        \midrule
        Pearson Correlation & 0.051& -0.018  \\
        Spearman Rank Correlation & 0.079  & -0.017  \\
        Kendall's Tau Correlation & 0.053  & -0.010  \\
        \bottomrule
    \end{tabular}
    }
    \caption{Correlation metrics between embedding similarities and the COCOLA score.}
    \label{tab:correlation}
\end{table}

The results show no meaningful correlation between either large audio model and the COCOLA score, with all correlation metrics yielding values close to zero. This suggests that the similarity measures computed from these embeddings fail to capture the musical coherence that COCOLA quantifies. While this outcome is consistent with the modality-based structure observed in CLAP’s embedding space (Section~\ref{sec:clustering}), it also reveals a limitation of MERT: despite being trained specifically for music, its embeddings appear unsuited for assessing mashup compatibility, at least in a zero-shot setting.

\section{Conclusion}
In this paper, we presented AutoMashup, a system for automatic mashup creation that combines recent advances in source separation, musical structure analysis, and compatibility estimation. Beyond the mashup generation process itself, we focused on the upstream task of selecting musically compatible tracks, and proposed the use of COCOLA to quantitatively assess stem compatibility in mashup creation. We also investigated whether general-purpose audio embeddings --- specifically CLAP and MERT --- could support this task in a zero-shot setting.

Our analysis revealed two key findings. First, mashup compatibility is inherently asymmetric: similarity between vocal and instrumental tracks depends on the direction of the pairing, highlighting the need to consider stem roles explicitly. Second, CLAP and MERT embeddings fail to align with perceptual musical compatibility, as measured by the COCOLA score. These results underscore the limitations of current general-purpose audio models in this creative musical context. 

Future work should explore alternative large audio models tailored to complex musical tasks, potentially by lifting the zero-shot constraint --- for instance, through COCOLA-based distillation or fine-tuning. Another promising direction is the development of representations sensitive to the functional roles of stems, such as by leveraging CLAP’s multimodal structure. Finally, perceptual evaluations of mashups should be conducted to assess their musicality. 

\bibliography{biblio}

\begin{thebibliography}{10}

\bibitem{brovig2012contextual}
Ragnhild Brøvig-Hanssen and Paul Harkins.
\newblock Contextual incongruity and musical congruity: the aesthetics and
  humour of mash-ups.
\newblock {\em Popular Music}, 31(1):87–104, 2012.

\bibitem{ciranni2025cocola}
Ruben Ciranni, Giorgio Mariani, Michele Mancusi, Emilian Postolache, Giorgio
  Fabbro, Emanuele Rodol{\`a}, and Luca Cosmo.
\newblock {COCOLA}: Coherence-oriented contrastive learning of musical audio
  representations.
\newblock In {\em ICASSP 2025}. IEEE, 2025.

\bibitem{davies2014automashupper}
Matthew~EP Davies, Philippe Hamel, Kazuyoshi Yoshii, and Masataka Goto.
\newblock Automashupper: Automatic creation of multi-song music mashups.
\newblock {\em IEEE/ACM Transactions on Audio, Speech, and Language
  Processing}, 22(12):1726--1737, 2014.

\bibitem{defferrard2017fma}
Michaël Defferrard, Kirell Benzi, Pierre Vandergheynst, and Xavier Bresson.
\newblock {FMA}: A dataset for music analysis.
\newblock {\em arXiv preprint arXiv:1612.01840}, 2017.

\bibitem{hooktheory2020analysis}
Hooktheory.
\newblock I analyzed the chords of 1300 popular songs for patterns.
\newblock Blog post, 2020.
\newblock
  \url{https://www.hooktheory.com/blog/i-analyzed-the-chords-of-1300-popular-songs}.

\bibitem{huang2021modeling}
Jiawen Huang, Ju-Chiang Wang, Jordan~BL Smith, Xuchen Song, and Yuxuan Wang.
\newblock Modeling the compatibility of stem tracks to generate music mashups.
\newblock In {\em Proceedings of the AAAI Conference on Artificial
  Intelligence}, volume~35, pages 187--195, 2021.

\bibitem{kim2023all}
Taejun Kim and Juhan Nam.
\newblock All-in-one metrical and functional structure analysis with
  neighborhood attentions on demixed audio.
\newblock In {\em IEEE Workshop on Applications of Signal Processing to Audio
  and Acoustics}. IEEE, 2023.

\bibitem{li2023mert}
Yizhi Li et~al.
\newblock Mert: Acoustic music understanding model with large-scale
  self-supervised training.
\newblock {\em arXiv preprint arXiv:2306.00107}, 2023.

\bibitem{pyrubberband}
Brian McFee.
\newblock Pyrubberband: Python wrapper for rubber band library.
\newblock \url{https://github.com/bmcfee/pyrubberband}.

\bibitem{popenstock2024mouth}
Popenstock.
\newblock Mouth souvenirs: La culture du mashup comme palimpseste du web.
\newblock {\em Popenstock}, 2024.

\bibitem{rand1971objective}
William~M Rand.
\newblock Objective criteria for the evaluation of clustering methods.
\newblock {\em Journal of the American Statistical Association},
  66(336):846--850, 1971.

\bibitem{richards2017tonal}
Mark Richards.
\newblock Tonal ambiguity in popular music's axis progressions.
\newblock {\em Music Theory Online}, 23(3), 2017.

\bibitem{rouard2023hybrid}
Simon Rouard, Francisco Massa, and Alexandre D{\'e}fossez.
\newblock Hybrid transformers for music source separation.
\newblock In {\em ICASSP 2023}. IEEE, 2023.

\bibitem{wong2023fma}
Stella Wong.
\newblock Key and mode annotations for the {FMA} dataset.
\newblock GitHub repository, 2023.
\newblock \url{https://github.com/stellaywong/fma_keys}.

\bibitem{wu2024graph}
Xinyang Wu and Andrew Horner.
\newblock Graph neural network guided music mashup generation.
\newblock In {\em 2024 IEEE International Conference on Big Data (BigData)},
  pages 3235--3241. IEEE Computer Society, 2024.

\bibitem{wu2023large}
Yusong Wu et~al.
\newblock Large-scale contrastive language-audio pretraining with feature
  fusion and keyword-to-caption augmentation.
\newblock In {\em ICASSP 2023}. IEEE, 2023.

\bibitem{xing2020pomash}
Baixi~andothers Xing.
\newblock Popmash: an automatic musical-mashup system using computation of
  musical and lyrical agreement for transitions.
\newblock {\em Multimedia Tools Appl.}, 79(29–30):21841–21871, 2020.

\end{thebibliography}

\end{document}